\documentclass[epj]{svjour}
\usepackage{latexsym}
\usepackage{graphicx}
\usepackage{cite}
\usepackage{epsfig}
\usepackage{amssymb}

\renewcommand{\abstractname}{Abstract}
\renewcommand{\refname}{\small References}         
\newcommand{\be}{\begin{equation}}                 
\newcommand{\ee}{\end{equation}}                   

\newcommand{\ega}{E_{\gamma}}
\newcommand{\cosp}{\cos\theta_\pi^*}
\newcommand{\cspi}{\cos\,\theta^*_{\pi}}
\newcommand{\agp}{A^p_{1/2}}
\newcommand{\agn}{A^n_{1/2}}
\newcommand{\bfeps}{\mbox {\boldmath $\epsilon$}}  
\newcommand{\bfe}{\mbox {\boldmath $e$}}           

\newcommand{\gdpid}{$\gamma d\!\to\!\pi^0 d$}      
\newcommand{\abra}{\left(\!\! \begin{array}{cc} }  
\newcommand{\aket}{\end{array}\!\!\right)}         
\newcommand{\NN}{$N(1535)S_{11}$}                  

\begin{document}
\hugehead
\title{Evidence for a backward peak in the
       \gdpid\ cross section near the $\eta$ threshold}

\author{Y.~Ilieva\inst{1}\fnmsep\thanks{On leave from: INRNE, BAS, Sofia, Bulgaria}\fnmsep\thanks{Current address: USC, Columbia, South Carolina 29208}
\and
B.L.~Berman\inst{1}
\and
A.E.~Kudryavtsev\inst{2,1}
\and
I.I.~Strakovsky\inst{1}
\and
V.E.~Tarasov\inst{2}
\and
M.~Amarian\inst{28}
\and
P.~Ambrozewicz\inst{14}
\and
M.~Anghinolfi\inst{19}
\and
G.~Asryan\inst{39}
\and
H.~Avakian\inst{33}
% ...
\newline(The CLAS collaboration)
}


\institute{
$^1$The George Washington University, Washington, District of Columbia 20052\\
$^2$Institute of Theoretical and Experimental Physics, Moscow, 117259, Russia\\
$^3$Argonne National Laboratory, Argonne, Illinois 60439\\
$^4$Arizona State University, Tempe, Arizona 85287\\
$^5$University of California at Los Angeles, Los Angeles, California  90095\\
$^6$California State University, Dominguez Hills, Carson, California 90747\\
$^7$Carnegie Mellon University, Pittsburgh, Pennsylvania 15213\\
$^8$Catholic University of America, Washington, District of Columbia 20064\\
$^9$CEA-Saclay, Service de Physique Nucl\'eaire, F91191 Gif-sur-Yvette, France\\
$^{10}$Christopher Newport University, Newport News, Virginia 23606\\
$^{11}$University of Connecticut, Storrs, Connecticut 06269\\
$^{12}$Edinburgh University, Edinburgh EH9 3JZ, United Kingdom\\
$^{13}$Fairfield University, Fairfield, Connecticut 06824\\
$^{14}$Florida International University, Miami, Florida 33199\\
$^{15}$Florida State University, Tallahassee, Florida 32306\\
$^{16}$University of Glasgow, Glasgow G12 8QQ, United Kingdom \\
$^{17}$Idaho State University, Pocatello, Idaho 83209 \\
$^{18}$INFN, Laboratori Nazionali di Frascati, 00044 Frascati, Italy \\
$^{19}$INFN, Sezione di Genova, 16146 Genova, Italy \\
$^{20}$Institut de Physique Nucleaire ORSAY, Orsay, France \\
$^{21}$James Madison University, Harrisonburg, Virginia 22807 \\
$^{22}$Kyungpook National University, Daegu 702-701, Republic of Korea \\
$^{23}$University of Massachusetts, Amherst, Massachusetts  01003 \\
$^{24}$Moscow State University, General Nuclear Physics Institute, 119899 Moscow, Russia\\
$^{25}$University of New Hampshire, Durham, New Hampshire 03824\\
$^{26}$Norfolk State University, Norfolk, Virginia 23504\\
$^{27}$Ohio University, Athens, Ohio  45701\\
$^{28}$Old Dominion University, Norfolk, Virginia 23529\\
$^{29}$Rensselaer Polytechnic Institute, Troy, New York 12180\\
$^{30}$Rice University, Houston, Texas 77005\\
$^{31}$University of Richmond, Richmond, Virginia 23173\\
$^{32}$University of South Carolina, Columbia, South Carolina 29208\\
$^{33}$Thomas Jefferson National Accelerator Facility, Newport News, Virginia 23606\\
$^{34}$TRIUMF, Vancouver, Canada V6T 2A3\\
$^{35}$Union College, Schenectady, NY 12308\\
$^{36}$Virginia Polytechnic Institute and State University, Blacksburg, Virginia 24061\\
$^{37}$University of Virginia, Charlottesville, Virginia 22901\\
$^{38}$College of William and Mary, Williamsburg, Virginia 23187\\
$^{39}$Yerevan Physics Institute, 375036 Yerevan, Armenia\\
$\dag$  deceased}

\date{Received: date / Revised version: date}

\abstract{
High-quality cross sections for the reaction \gdpid\ have been
measured using the CLAS at Jefferson Lab over a wide energy range near
and above the $\eta$-meson photoproduction threshold.
At backward c.m. angles for the outgoing pions, we observe a resonance-like
structure near E$_{\gamma}$=700~MeV. Our model analysis shows that it can
be explained by $\eta$
excitation in the intermediate state. The effect is the result of
the contribution of the \NN\ resonance to the amplitudes of the
subprocesses occurring between the two nucleons and of a two-step
process in which the excitation of an intermediate $\eta$ meson
dominates.
} 
%
\maketitle
\markboth{The CLAS Collaboration (Y. Ilieva {\it et al.)}: A backward peak in the
       \gdpid\ cross section near the $\eta$ threshold}{}
\section{Introduction}
\label{intro}
Interactions of the $\eta$ meson with few-nucleon systems
complement our knowledge of the  $\eta$-nucleon interaction.
Interest in these systems stems from the hypothetical
existence of $\eta$-nuclear quasibound states. Such states have
been predicted by Haider and Liu~\cite{Haider} and Li, Cheung, and
Kuo~\cite{Li}.  Although there has been no direct
experimental verification of this hypothesis to date, there is mounting
evidence that such states might exist in the lightest few-nucleon
systems \cite{Mayer,Willis,Calen}.

For the case of the three-nucleon system, it
was found at Saclay that the \mbox{$dp\to\eta \,{\rm ^3He}$} production amplitude
falls rapidly just above the $\eta$ threshold~\cite{Mayer}.
A less pronounced slope was found in the \mbox{$dd\rightarrow \eta \,{\rm ^4He}$}
amplitude~\cite{Willis}.
For the two-nucleon system, very strong final-state
interactions (FSI) were found in the $pp\to pp\eta$ cross
section in the threshold region~\cite{Calen}.  The
energy dependence of the $NN\to NN\eta$ and $NN\to
d\eta$ reactions can be understood in terms of $NN$ FSI (\cite{Baru}
and references therein).
However, as noted
in Ref.~\cite{Green}, the existence of a narrow
virtual state in the $\eta$-deuteron system can be inferred.

The production of a virtual $\eta$ meson may also play a role in
other nuclear reactions, even in those for which there is no $\eta$
in either the initial or
the final state, but only in an intermediate state.
Examples are the reactions \mbox{$pd\to\pi^+ \,{\rm ^3H}$} and \mbox{$pd\to{\pi^0} \,{^3{\rm He}}$},
for which there are strong indications that an intermediate
$\eta$ cusp is present~\cite{Mach}.
Evidence for an intermediate $\eta$ meson also was found
in elastic
$\pi d$ backward scattering. This contribution manifests itself
as a cusp in the energy dependence of the
backward differential cross sections near the $\eta$ threshold.
The effect was predicted in Ref.~\cite{Kondr} and was confirmed
by several independent measurements of backward $\pi
d$ scattering~\cite{Abram}.  In this work, we present
the first systematic evidence for a similar phenomenon
(the first indication was found in Ref.~\cite{imanishi})
in the coherent photoproduction of a neutral pion on the deuteron,
\gdpid .

All of the above phenomena take place because
$\eta$-meson production in hadron-hadron collisions
near threshold is enhanced.  This is because the
cross section for excitation of the
nearby baryonic resonance \NN\ is large and
this resonance is strongly coupled to the $\eta N$ channel.
Since the amplitudes for
photoproduction of the \NN\ are also large~\cite{Arndt,Arndt90,Knoch},
one can expect a similar enhancement
in various photonuclear reactions.

\begin{figure}[ht]
\centerline{\includegraphics[height=0.13\textwidth,angle=0]{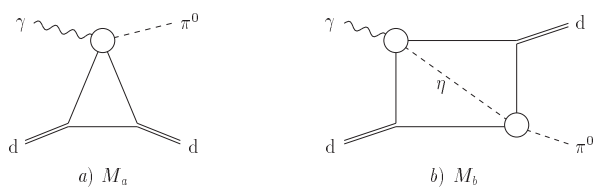}}
\caption{Feynman diagrams for the \gdpid\ reaction considered in~\cite{photo}: (a) single-scattering amplitude $M_a$;
         (b) double-scattering amplitude $M_b$. It was shown in~\cite{photo} that
(b) dominates over (a) at backward angles for $E_\gamma\sim 700$ MeV. \label{fig:g1}}
\end{figure}

Recently, coherent
photoproduction of the $\pi^0$ meson from the deuteron was studied
theoretically \cite{photo}. In particular, it was demonstrated that at large
c.m. scattering angles and photon energies $E_{\gamma}$ between 600 and 800 MeV,
the two-step process with the
excitation of an intermediate $\eta$-meson (shown in Fig.~\ref{fig:g1}(b)) dominates over single-step
photoproduction (shown in Fig.~\ref{fig:g1}(a)) and pion rescattering. This two-step
process can be analyzed as
two sequential subprocesses:
$\gamma N_1\to N(1535)\to\eta N_1$ and
$\eta N_2\to N(1535)\to\pi^0N_2$, where
$N_1$ and $N_2$ are the two nucleons in the deuteron.
It was shown in~\cite{photo} that this
mechanism explains qualitatively the structure in the \gdpid\ differential cross section,
which we present here, at large angles and for
$E_{\gamma}\sim$ 600--800~MeV.  The main conclusions of Ref.~\cite{photo}
were reproduced in a more recent paper~\cite{Fix}, where
it was shown that in addition to this two-step process,
the full dynamics in the intermediate $NN\eta$ system could be important as well.
Other theoretical studies of the \gdpid\ reaction
can be found in a number of papers~\cite{Tez}. However, none
of these considers the effect of the opening of the $\eta$ threshold
at 700 MeV.

Our photoproduction data, presented here, give
for the first time clear evidence for a prominent effect around 700 MeV
at large c.m. angles, which can be explained by the excitation of an intermediate
$\eta$ meson.

\section{Experiment}
\label{experiment}
The experiment took place in Hall B at Jefferson Lab using the
CEBAF Large Acceptance Spectrometer (CLAS)~\cite{clas}.  A collimated, tagged, real-photon beam, with
energies between 0.5 and 2.3~GeV was incident on a 10-cm-long
LD$_2$ target installed in the center of the CLAS. The energies of
the photons were tagged using the Hall-B tagger~\cite{tag}.
The outgoing deuterons were tracked in the six toroidal magnetic
spectrometers of the CLAS. They were bent outwards by the
magnetic field, and their trajectories were measured by three
layers of drift chambers surrounding the LD$_2$ target. The
time of flight of the deuterons was measured by 6$\times$48 scintillators (TOF)
that surround the CLAS detector outside of the magnetic field.
A set of six
scintillator counters, comprising the start counter and placed
just around the target, measured the event time at the
vertex.  The CLAS covers the polar angular range
between 8$^\circ$ and 142$^\circ$ in the laboratory system and the entire
range in azimuthal angle.

\section{Data Analysis}
\label{dataanalysis}
Since the CLAS has very good acceptance for charged particles,
and a very limited one for neutrals, the analysis of the \gdpid\
reaction was based on detecting the final-state deuterons and
selecting the good events by the missing-mass technique. Deuterons
were identified from their time of flight and momentum, which
allowed a mass reconstruction. Thus, the initial event selection was done based
on the reconstructed mass.
Further selection of events was achieved by
comparing the event vertex time with the photon vertex time as
measured by the tagger. Low-momentum protons were discarded based
on a cut on the particle momentum \textit{vs.} energy loss in the TOF. In
addition, fiducial cuts were applied to the remaining data sample
in order to remove edge areas of the detector where the acceptance
was not well reproduced by a simulation.

\begin{figure}
\centerline{
\includegraphics[height=0.24\textwidth, angle=0]{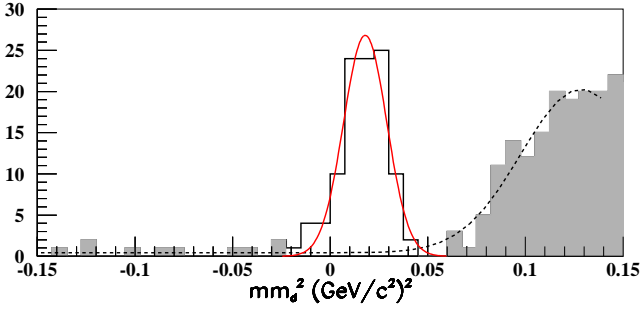}}
\caption{An example of the background subtraction, for the bin
         $0.675<E_\gamma<0.7\textrm{ GeV}$ and
         $-0.8<\cosp < -0.7$. One sees that the pion peak is well separated from
the background. The contribution of the latter to the peak is $< 10\%$.
The shaded areas show the sidebands around the pion peak we used in order to
determine the shape and the magnitude of the background underneath it.
\label{fig:g2}}
\end{figure}

Once the data were reduced, based on all of the above cuts,
they were binned in photon energy and pion c.m. scattering angle
$(E_\gamma, \cosp)$.  Here, we present
differential cross sections for the reaction \gdpid\ based on
a photon-energy bin width of 25~MeV and a $\cosp$
bin width of 0.1. For every such $(E_\gamma, \cosp)$
bin, we determined the reaction yields after sideband background
subtraction was performed on the missing-mass ($mm^2_d$)
distributions as shown in Fig.~\ref{fig:g2}. The quantity $mm^2_d$ is
defined as
$mm^2_d=(p_{\gamma}+ p_t - p_d)^2$,
where $p_\gamma$, $p_t$, and $p_d$ are the four-momentum vectors
of the beam, the target, and the recoil deuteron, respectively.
The systematic uncertainty associated with the background
subtraction depends on the kinematic bin, and varies between
1.5$\%$ and 6.4$\%$. The statistical uncertainty of the extracted
yields is also bin-dependent, and varies between 3$\%$ and
30$\%$.

\section{Results}
\label{results}
Differential cross sections were obtained by normalizing the true-event
yields
to the photon flux, the number of target scattering centers, and
the CLAS acceptance \cite{anote}. The statistical uncertainty of the photon
flux is negligible. The systematic uncertainty of the evaluated
photon flux is 3.3$\%$. The statistical uncertainty associated
with the calculated CLAS acceptance is bin-dependent and varies
between 0.8$\%$ and 2.5$\%$. The systematic uncertainty of the
acceptance is less than 10\%.
A common factor of $1.0141\pm 0.0006^{stat}\pm 0.0005^{syst}$,
due to the inefficiency of the procedure for choosing the right
photon for a given event, was applied to all of the differential
cross sections. There is also an overall systematic uncertainty of
0.5$\%$ related to the determination of the target length and
density~\cite{rossi}. Thus, the total uncertainty of the differential
cross sections is bin-dependent and varies smoothly from 11$\%$ to 33$\%$.

\begin{figure}
\centering
\includegraphics[height=0.45\textwidth, angle=90]{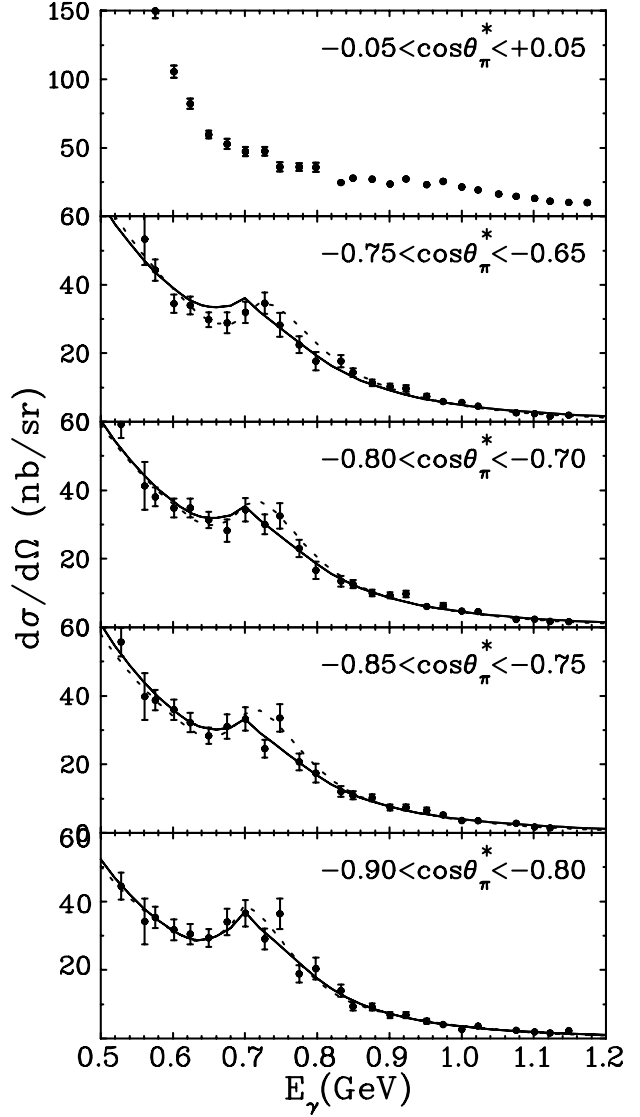}
\caption{Excitation functions for the reaction $\gamma d\rightarrow \pi^0 d$ for
several bins in $\cosp$.
The filled circles are our CLAS data.
The error bars show the total bin-dependent uncertainty for each data point.
The dashed and solid curves are the results of fit 1 and fit 2
(see the text), respectively, obtained with the helicity
amplitudes from Ref.~\cite{Arndt90} (set (2)).
         One sees that the excitation function at $\cosp=0.0$ does
         not exhibit any prominent structure near 0.7~GeV. Note that
some of the angular bins overlap partially with each other.
\label{fig:g3}}
\end{figure}

Our differential cross sections for
the reaction \gdpid\ for five c.m. angular bins are shown in
Fig.~\ref{fig:g3}. Note that some of
the angular bins overlap partially with each other.
The latter is due to the fact
that for a consistency check, we determined differential cross sections
for two entirely separate angular binnings.
The figure illustrates the consistency of the structure
and the model interpretation at different kinematical binnings.

Overall, our data are in very
good agreement with
previously measured differential
cross sections \cite{imanishi,baba,meekins,asai}.
However, in the range of photon energies where we observe the backward peak
discussed here, the data of Ref.~\cite{imanishi} exhibit a structure but of a smaller
magnitude,
whereas the data of Ref.~\cite{asai} do not exhibit any structure.
In order to understand this discrepancy we performed many
consistency checks, and our analysis procedure was studied
in great detail for possible sources of errors~\cite{anote}. The structure near
0.7 GeV and its magnitude persists.
Both of the measurements \cite{imanishi,asai} were done at the same
accelerator facility and used an untagged bremsstrahlung beam, which might
introduce large systematic uncertainties in the determination of the photon flux.

\section{Discussion}
\label{discussion}
To achieve a quantitative understanding of our data,
we employ the semiphenomenological description of Ref.~\cite{photo}.
We express
the reaction amplitude $M$ as a sum $M\!=M_1 S+\!M_2 S$, where
$M_1 S$ is the two-step amplitude $M_b$ given by the diagram
in Fig.~\ref{fig:g1}(b) and calculated in Ref.~\cite{photo}
with intermediate $\eta$ production, and $S$ is a spin factor.
$M_2 S$ is the
effective ``background" amplitude which takes into account
all possible background diagrams (including the single-scattering
amplitude $M_a$ shown in Fig.~\ref{fig:g1}(a)).
We parametrize it as $M^{}_2=A\exp(i\varphi_2-b \ega)$, where
$\varphi_2$ is the phase. The square of the total amplitude with
unpolarized particles is then written as
$|M|^2=|M_1+M_2|^2 \overline{|S|^2}$ and is applied to describe
the experimental excitation functions for given values of
$\cspi$. We give the results of fits for two different
parametrizations of $\varphi_2$:
\begin{eqnarray}
{\rm fit~1:}~~~\varphi_2(\ega)&=&\alpha+\beta\,(\ega-\!0.7\,\textrm{GeV});
\label{1} \\
{\rm fit~2:}~~~\varphi_2(\ega)&=&\alpha+\varphi_1(\ega).~~~~~~~~~~~~~~
\label{2}
\end{eqnarray}
For fit~1, we use a linear $\ega$-dependent background phase
$\varphi_2(\ega)$ with two parameters, $\alpha$ and $\beta$. We define
$\alpha=\varphi_2(\ega)$ at energy $\ega=0.7$ GeV, where
``by eye" the amplitude $M_1$ peaks. Thus, we use four parameters
($A$, $b$, $\alpha$, and $\beta$) in fit~1.

For fit~2, we use the parameter $\alpha$ as the relative
phase of the amplitudes $M_1$ and $M_2$, \textit{i.e.},
$\alpha=\varphi_2(\ega)\!-\!\varphi_1(\ega)$, where $\varphi_1(\ega)$
is the phase of the amplitude $M_1$ and is a given function. Thus,
we use three parameters ($A$, $b$, and $\alpha$) in fit~2.
In both variants all the parameters were varied
independently to fit the data for various values of $\cspi$.

We note that the two-step amplitude $M_1$ with intermediate
$\eta$ production is proportional to $\agp\!-\!\agn$, where
$\agp$ ($\agn$) is the helicity amplitude of the decay
$N(1535)\!\to\!p\gamma(n\gamma)$. Values for these amplitudes
vary widely in the literature. Here, we give some sets
(in units of 10$^{-3}\,$GeV$^{-1/2}$):

(1) $\agp=107$,~$\agn=-96$~(\cite{Knoch});

(2) $\agp=~78$,~$\agn=-50$~(\cite{Arndt90}, used in \cite{photo});

(3) $\agp=~60$,~$\agn=-20$~(\cite{Arndt96}).

The results for the free parameters from the data fitting
are given in Table~\ref{tab:table1} for large values of $\theta^*_{\pi}$.
Here, the coupling set (2) was used.
The dashed and the solid curves in Fig.~3 correspond to
the results of fits~1 and 2, respectively, and both
fits give a satisfactory description of the data.
The corresponding values of $\chi^2/N$ ($N$ is the number of
degrees of freedom) are also given in Table~\ref{tab:table1}.
\begin{table*}[!hbt]
\caption{\label{tab:table1}Parameters and $\chi^2$ per degree of freedom $N$ of the
model-based fits, explained  in the text, to the experimental
excitation functions at large c.m. angles. Parameters $A$, $b$,
$\alpha$, and $\beta$ ($A$, $b$, and $\alpha$) were used in
fit~1 (fit~2). The values for the helicity amplitudes $A^{p,n}_{1/2}$
used in the calculations were taken from Ref.~[12] (set~(2)).
$N=N_{dat}-N_{par}$, where $N_{par}=4(3)$ for fit~1(2).
$N_{dat}=24$ for $\cspi\!=\!-0.8$, and 25 for the others.}
\begin{center} \begin{tabular*}{0.9\textwidth}{@{\extracolsep{\fill}}ccccccc}
\hline
$\cspi$ & fit & $A$ & $b$~(GeV$^{-1}$) & $\alpha$~(deg) & $\beta$~(deg/MeV) & $\chi^2/N$ \\
\hline
-0.7  & 1 & 1.68 & 2.53 & 177 & 0.79 & 1.12 \\
-0.7  & 2 & 1.55 & 2.40 &  59 &      & 1.73 \\
-0.75 & 1 & 1.57 & 2.51 & 156 & 0.80 & 1.27 \\
-0.75 & 2 & 1.52 & 2.42 &  59 &      & 1.07 \\
-0.8  & 1 & 1.69 & 2.68 & 158 & 0.81 & 1.55 \\
-0.8  & 2 & 1.63 & 2.55 &  66 &      & 0.83 \\
-0.85 & 1 & 1.51 & 2.59 & 136 & 0.89 & 0.99 \\
-0.85 & 2 & 1.55 & 2.61 &  42 &      & 0.79 \\
\hline
\end{tabular*} \end{center}
\end{table*}
Using this procedure, we obtain a good description of the data at
large scattering angles ($\cspi< -0.5$), where the $\eta$ effect
is strongly pronounced. We do not consider the data at smaller
angles because a more complicated parametrization of the background
is needed and in any case the effect is less pronounced.

The backward-angle structure is qualitatively reproduced using
various sets of couplings $\agp$ and $\agn$. We obtained also
a good description (not shown) using coupling set (1) in fit~2
and set (3) in fits~1 and 2.
Thus, the experimentally observed enhancement in the excitation
functions near $\ega\sim 0.7$~GeV needs no exotic explanation
such as was done, for instance, in Ref.~\cite{imanishi}.

In order to evaluate the statistical significance of the structure
exhibited by our data, we performed a $\chi^2$ test of goodness of fit. We tested the hypothesis
that the excitation functions at large angles, which are shown in Fig. \ref{fig:g3}, can be described by
a smooth function. We used the amplitude $M_2$, which falls off smoothly with $E_\gamma$ and
describes well the experimental data outside the energy range where the enhancement is observed,
to fit the excitation functions at large angles. The results for the $\chi^2$ and the significance
level of this hypothesis (the P-value) are given in Table \ref{tab:table2}.
\begin{table}[!ht]
\caption{\label{tab:table2}$\chi^2$ and number of degrees of freedom $N$ of the
fits to the large angle excitation functions with the smooth amplitude $M_2$. The significance
of the hypothesis (the P-value for the $\chi^2$ test of goodness of fit)
that $M_2$ describes the data is also shown.}
\begin{center} \begin{tabular*}{0.45\textwidth}{@{\extracolsep{\fill}}cccc}
\hline  $\cspi$ & $\chi^2$ & $N$ & P-value  \\
\hline
-0.7  & 44.43 & 22    & 0.003  \\
-0.75 & 43.68 & 22    & 0.004 \\
-0.8  & 35.71 & 21    & 0.02  \\
-0.85 & 77.07 & 22    & 0.0001 \\
\hline
\end{tabular*} \end{center}
\end{table}
We see that the probability that this hypothesis is true is smaller than 2$\%$. Thus,
the statistical significance of the structure is greater than 98$\%$.

Finally, the excitation of the \NN\ resonance
depends on the initial polarization of the beam and target.
The spin factor $S$ of the two-step amplitude with an
intermediate $\eta$ meson and
an $S$-wave deuteron wave function
is $S=(\bfeps^*_2\cdot
[\bfe\times\bfeps_1])$, where $\bfe$ and $\bfeps_{1}$ ($\bfeps_2$)
are the polarization 3-vectors of the initial photon and the initial
(final) deuteron. For an unpolarized final-state
deuteron and for $\lambda_{\gamma}=\pm 1$ and $\lambda_d=0,\pm 1$
as the helicities of the initial photon and deuteron,
respectively, we obtain $|S|^2=2/3$ for unpolarized
particles; $|S|^2=1$ for $\lambda_d=0,-\lambda_{\gamma\,}$; and
$|S|^2=0$ for $\lambda_d=\lambda_{\gamma\,}$. Thus, for
$\lambda_d=0,-\lambda_{\gamma\,}$, we expect the $\eta$-effect
to be enhanced compared with the unpolarized case.
For $\lambda_d=\lambda_{\gamma\,}$ (parallel polarization of
the initial photon and deuteron), the excitation of the $N(1535)S_{11}$,
and hence the
$\eta$-effect, should be suppressed.

\section{Summary}
\label{summary}
To summarize, we have measured unpolarized differential cross sections for the \gdpid\ reaction
at backward c.m. angles and for photon
energies above 500 MeV. The data show
a pronounced structure (with statistical significance greater than 98$\%$)
in the excitation functions in the region of 700 MeV.
For the first time, this phenomenon is systematically studied with good accuracy.
The structure can be explained by the opening of the
$\eta$-photoproduction threshold on a single nucleon, and in our model is mainly related to the
excitation of the intermediate resonance \NN\ in a two-step
process.
The details of the underlying dynamics
can be further explored via polarization measurements.

\section*{Acknowledgement}
We would like to acknowledge the efforts of the staff of the
Accelerator and the Physics Divisions at Jefferson Lab that made this experiment possible.
This work was supported by the U.~S.~Department of Energy under grant
DE--FG02--95ER40901,
in part under grant DE--FG02--99ER41110, by the grant of the Russian
Ministry of Industry, Science, and Technology NSh 5603.2006.2, by the
Science and Technology Facilities Council (STFC), and by the National Research Foundation of Korea.
The Southeastern Universities Research Association (SURA) operated the
Thomas Jefferson National Accelerator Facility for the United States
Department of Energy under contract DE--AC05--84ER40150 until May 31, 2006.

\end{document}